\documentclass[showpacs,prl,aps,twocolumn]{revtex4-1}
\usepackage{amsmath,amsfonts,amssymb,bm,graphicx}

\newcommand{\WTD}{\mathcal{W}}
\newcommand{\W}{\mathbb{W}}
\newcommand{\rhohat}{\hat{\rho}}
\newcommand{\gamhat}{\hat{\gamma}}
\newcommand{\G}{\mathcal{G}}

\newcommand{\hatH}{\hat{H}}

\begin{document}
\title{Electron Waiting Times in Non-Markovian Quantum Transport}

\author{Konrad H.\ Thomas}
\author{Christian Flindt}
\affiliation{D\'epartement de Physique Th\'eorique, Universit\'e de Gen\`eve, 1211 Gen\`eve, Switzerland}
\date{\today}

\begin{abstract}
We formulate a quantum theory of electron waiting time distributions for charge transport in nano-structures described by non-Markovian generalized master equations. We illustrate our method by calculating the waiting time distribution of electron transport through a dissipative double quantum dot, where memory effects are present due to a strongly coupled heat bath. We consider the influence of non-Markovian dephasing on the distribution of electron waiting times and discuss how spectral properties of the heat bath may be detected through measurements of the electron waiting time.
\end{abstract}



\maketitle

\emph{Introduction}.--- The theory of open quantum systems is important in many branches of physics \cite{Breuer02}. The typical scenario consists of a small quantum system with a few degrees of freedom coupled to a large environment. The system evolves coherently due to its internal dynamics, but also undergoes non-unitary evolution as it interacts with the environment. For weak system-environment couplings, the dynamics of the quantum system is often Markovian. However, as the coupling increases, information about the system, which leaks out into the environment, may flow back to the system at a later time, making the evolution non-Markovian. Deterministic control of the transition from Markovian to non-Markovian dynamics of an open quantum system was recently demonstrated in a quantum optical experiment \cite{Liu11}.

\begin{figure}
\includegraphics[width=0.85\columnwidth]{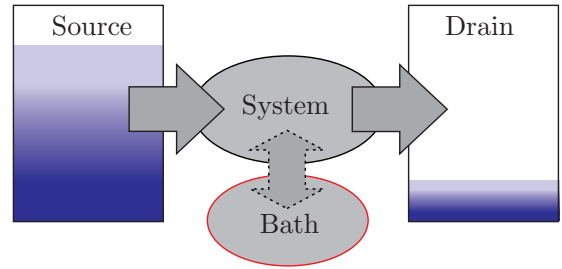}
\caption{(color online). Generic quantum transport setup. A nano-scale electronic system is connected to source and drain electrodes as well as an external heat bath. The applied bias drives electrons through the system from the source to the drain electrode. The system exchanges energy with the bath, but no particles are transferred between system and bath. We consider situations where the dynamics of the system is non-Markovian due to the electronic reservoirs and the heat bath. \label{Fig1}}
\end{figure}

In electronic transport, the theory of open quantum systems can describe non-equilibrium charge flow through nano-electronic conductors. The conductor exchanges  particles with the external electronic leads and may also interact with a heat bath, Fig.~\ref{Fig1}. The full counting statistics of transferred charges \cite{Braggio06,Flindt08,Flindt10,Emary09a,Zedler09,*Zedler11,Emary11a} as well as the finite-frequency current noise \cite{Aguado04a,*Aguado04b,Engel04,Schaller09,Marcos11,Emary11b} have been investigated intensively for non-Markovian transport processes, and shot noise measurements have revealed strong memory effects in the transport through a quantum dot in resonance with the Fermi level of an external electrode~\cite{Ubbelohde12b}.

A very recent interest in quantum transport concerns the distribution of waiting times between consecutive charge transfers \cite{Brandes08,Welack09b,Albert11,Albert12}. This line of research seems particularly relevant in the light of the increasing number of accurate single-electron counting experiments \cite{Bylander05,*Fujisawa06,*Gustavsson06,*Flindt09,*Gustavsson09,*Maisi11,*Ubbelohde12a}. Theories have now been developed to describe electronic waiting time distributions (WTDs) for driven single-electron emitters \cite{Albert11} and phase-coherent conductors \cite{Albert12}. In quantum optics, non-Markovian effects in the decay dynamics of laser-driven systems have been examined \cite{Budini06} and stochastic simulations are currently being used to extract the WTDs of non-Markovian quantum systems \cite{Luoma12}. In electronic transport, Brandes has developed a compact and elegant method to calculate WTDs for systems described by Markovian generalized master equations (GMEs)~\cite{Brandes08}.

In this Letter we consider quantum transport governed by a generic non-Markovian GME and derive a general expression for the electronic WTD in Laplace space. The GME contains the memory kernel of the transport process and an inhomogeneity term, which accounts for memory effects at the time, when charge detection begins. This term has been the subject of recent theoretical investigations \cite{Emary11b,Marcos11}. It typically decays on the same time scale as the memory kernel and is not important for the long-time limit of time-integrated quantities like the zero-frequency noise and higher cumulants of the current, but is crucial to include when investigating correlation functions and fluctuations at finite times and frequencies \cite{Flindt08,Emary11b,Marcos11}. As we demonstrate below, it must also be incorporated together with the kernel in a consistent theory of WTDs for non-Markovian quantum transport.

We illustrate our methodology by evaluating the WTD for a dissipative double quantum dot (DQD) whose dynamics is non-Markovian due to a strongly coupled heat bath. This electronic analogue of an open spin-boson problem provides us with a microscopic model that can describe the transition from Markovian to non-Markovian dephasing. We show how coherent oscillations between the quantum dots are washed out by an increasing bath temperature, which dephases electrons as they propagate through the DQD, and we discuss WTDs as the system-bath coupling becomes strong. We then tune the DQD to a parameter regime, where the WTD becomes particularly sensitive to the spectral properties of the heat bath, in a  similar spirit to proposals for detecting the high-frequency quantum noise of a mesoscopic conductor by measuring the noise-induced inelastic current in a nearby DQD \cite{Aguado00}. Within this approach we  demonstrate how the absorption and emission of energy quanta to and from the heat bath can be clearly identified in the electronic WTD.

\emph{Non-Markovian GME.}--- We consider a generic non-Markovian GME of the form \cite{Flindt08,Flindt10}
\begin{equation}
\frac{d}{dt}\rhohat(n,t)\!=\!\!\sum_{n'=0}^{\infty}\int_0^{t}\!\!dt'\, \W(n\!-\!n',t\!-\!t')\rhohat(n',t')+\gamhat(n,t),
\label{eq:GME}
\end{equation}
describing charge transport through a nano-scale conductor as illustrated in Fig.~\ref{Fig1}. Here $\rhohat(n,t)$ is the reduced density matrix of the quantum system, obtained by tracing out the external electronic reservoirs and the heat bath. It has been resolved with respect to the number of transferred electrons, such that $P(n,t)=\mathrm{Tr}\{\rhohat(n,t)\}$ is the probability of having collected $n$ electrons in the drain during the time span $[0,t]$ \cite{Makhlin2001}. We assume low electronic temperatures compared to the applied voltage so that thermal charge fluctuations are negligible and we may focus on the uni-directional non-equilibrium charge transport from source to drain with $n$ being non-negative.

The kernel $\W(n,t)$ determines the time evolution of $\rhohat(n,t)$, taking into account memory effects due to the electronic reservoirs and the external heat bath.  The inhomogeneity $\gamhat(n,t)$ describes memory effects from before $t=0$, when charge detection begins. The first term on the right hand side of Eq.~(\ref{eq:GME}) does not include such memory effects. Non-Markovian GMEs as Eq.\ (1) arise in a variety of contexts, for example in the real-time diagrammatic technique \cite{Schoeller94,*Konig96,*Konig96b} and in the Nakajima--Zwanzig projection method \cite{Zwanzig01}. They have recently been investigated in connection with full counting statistics \cite{Braggio06,Flindt08,Emary09a,Flindt10,Zedler09,*Zedler11,Flindt10,Emary11a} and finite-frequency noise \cite{Flindt08,Marcos11,Emary11b}. As we go on to show, they also provide a useful starting point for calculating WTDs in non-Markovian quantum transport.

\emph{Distribution of electron waiting times.}--- To find the WTD, we first solve Eq.~(\ref{eq:GME}) for $\rhohat(n,t)$. To this end, we employ an operator-valued generalization of the generating function technique in Laplace space by introducing the transformed reduced density operator $\tilde{\rhohat}(s,z)=\sum_{n=0}^{\infty}\int_0^\infty dt \rhohat(n,t) s^n e^{-zt}$ and similarly for $\widetilde{\W}(s,z)$ and $\tilde{\gamhat}(s,z)$ \cite{Note1}. Within this framework, the reduced density matrix is readily obtained as $\tilde{\rhohat}(s,z)=\widetilde{\G}(s,z)\{\tilde{\gamhat}(s,z)+\rhohat^{\mathrm{stat}}\}$, where $\widetilde{\G}(s,z)=[z-\widetilde{\W}(s,z)]^{-1}$ is the resolvent of the kernel and $\rhohat^{\mathrm{stat}}$ is the stationary state. In the distant past, the system is prepared in an arbitrary state, but the stationary state is reached before detection of charges begins at $t=0$. The stationary state satisfies $\widetilde{\W}(1,0)\rhohat^{\mathrm{stat}}=0$ with the normalization $\mathrm{Tr}\{\rhohat^{\mathrm{stat}}\}=1$.

For stationary processes, the WTD is related to the idle time probability $\Pi(\tau)$ as $\WTD(\tau)=\langle\tau\rangle\partial^2_\tau\Pi(\tau)$, where the mean waiting time is determined by the average particle current as $\langle\tau\rangle=1/\langle I\rangle$ \cite{Vyas88,Albert12}. Here $\Pi(\tau)$ is the probability of not observing any electrons in a time interval of length $\tau$, $[t_0,t_0+\tau]$. This probability is independent of $t_0$ and we may take $t_0=0$. The average particle current is well-known and reads $\langle I\rangle=\mathrm{Tr}\{\partial_s\widetilde{\W}(s,0)\rhohat^{\mathrm{stat}}\}¦_{s=1}$ \cite{Braggio06,Flindt08,Flindt10}. In Laplace space, the WTD becomes $\widetilde{\WTD}(z)=\langle\tau\rangle z [z\widetilde{\Pi}(z)-1]+1$, where $\Pi(\tau=0)=1$ and $\partial_\tau\Pi(\tau=0)=-1/\langle\tau\rangle$ have been used. Importantly, the idle time probability can be expressed in terms of the reduced density matrix as $\widetilde{\Pi}(z)=\widetilde{P}(n=0,z)=\mathrm{Tr}\{\tilde{\rhohat}(s=0,z)\}$. Inserting the solution for $\tilde{\rhohat}(s,z)$, we arrive at the key result of this section
\begin{equation}
\widetilde{\WTD}(z)=\langle\tau\rangle z\left[z\langle \widetilde{\G}(z)\rangle-1\right]+1
\label{eq:WTDlap}
\end{equation}
with $\langle \widetilde{\G}(z)\rangle=\mathrm{Tr}\left[\widetilde{\G}(0,z)\{\tilde{\gamhat}(0,z)+\rhohat^{\mathrm{stat}}\}\right]$. This equation generalizes the expression by Brandes for Markovian GMEs \cite{Brandes08}. It is easy to show that our result reduces to that of Ref.~\cite{Brandes08}, if Eq.~(\ref{eq:GME}) is replaced by a Markovian GME. We see that the WTD corresponding to a generic non-Markovian GME includes both the kernel and the inhomogeneity $\tilde{\gamhat}$ \cite{Note2}. To revert the WTD to the time domain an inverse Laplace transformation must be performed analytically or numerically as illustrated below.

\emph{Dissipative DQD.}--- As a concrete application of our method, we consider charge transport through a DQD, where non-Markovian effects occur due to a strongly coupled heat bath. The total Hamiltonian of the setup reads
\begin{equation}
\hatH = \hatH_S+\hatH_T+\hatH_L+\hatH_{SB}+\hatH_B,
\label{eq:DQDham}
\end{equation}
where $\hatH_S=(\varepsilon/2)(\hat{d}_L^{\dag}\hat{d}_L-\hat{d}_R^{\dag}\hat{d}_R)+T_c(\hat{d}_L^{\dag}\hat{d}_R+\hat{d}_R^{\dag}\hat{d}_L)$ describes the left and right levels of the DQD with dealignment $\varepsilon$ and tunnel-coupling $T_c$; tunneling between left (right) level and left (right) lead is accounted for by $\hatH_T=\sum_{k,\alpha=L,R}t_{k,\alpha}\hat{c}_{k,\alpha}^{\dag}\hat{d}_\alpha+\mathrm{h.\,c.}$; the Hamiltonian of the leads is $\hatH_L=\sum_{k,\alpha=L,R}\epsilon_{k,\alpha}\hat{c}_{k,\alpha}^{\dag}\hat{c}_{k,\alpha}$; the coupling between the DQD and the heat bath reads $\hatH_{SB}=(\hat{d}_L^{\dag}\hat{d}_L-\hat{d}_R^{\dag}\hat{d}_R)\sum_j(g_j/2)(\hat{a}_j^{\dag}+\hat{a}_j)$; and $\hatH_B=\sum_j\hbar\omega_j\hat{a}_j^{\dag}\hat{a}_j$ describes the heat bath as an ensemble of harmonic oscillators. This is an open spin-boson problem, where the r\^{o}le of the spin is played by the two single-particle levels of the DQD which are coupled to a bath of bosons. It is a transport problem as charges enter and leave the pseudo-spin states from the voltage-biased electrodes.

\begin{figure*}
\includegraphics[width=0.85\textwidth]{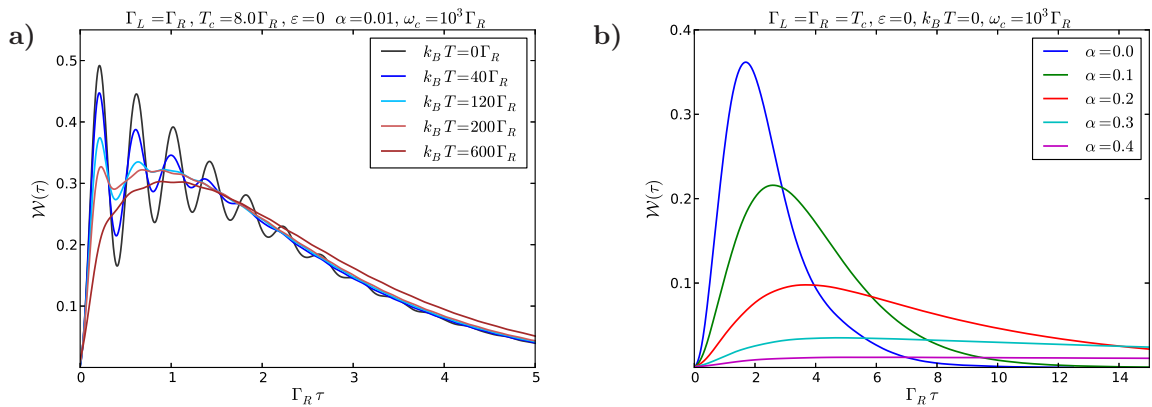}
\caption{(color online). Electronic waiting time distributions for a dissipative double quantum dot. (a) Coherent oscillations between the quantum dots are visible for weak couplings to the heat bath ($\alpha=0.01$). The electrons are dephased by an increasing bath temperature which washes out the coherent oscillations. (b) Electronic waiting time distributions beyond the weak-coupling limit. The increased coupling to the heat bath tends to localize electrons on the quantum dots.}
\label{Fig2}
\end{figure*}

We derive a non-Markovian GME for the populations of the DQD by tracing out the electronic leads and the heat bath \cite{Note3}. The reduced density matrix $\rhohat(n,t)=[\rhohat_{0}(n,t),\rhohat_{L}(n,t),\rhohat_{R}(n,t),\rhohat_{D}(n,t)]^T$ contains the probabilities for the DQD to be empty, having left or right dot occupied, or to be doubly occupied.  In Laplace space, the kernel reads
\begin{equation}
\widetilde{\W}(s,z)\!=\!\! \left(
                       \begin{array}{cccc}
                         -\Gamma_L\! & 0\! & s\Gamma_R\! & 0\! \\
                         \Gamma_L\! & -\widetilde{\Gamma}_{\!+}(z)\! & \widetilde{\Gamma}_{\!\!-}(z)\! & s\Gamma_R\! \\
                         0\! & \widetilde{\Gamma}_{\!+}(z)\! & -\widetilde{\Gamma}_{\!\!-}(z)\!-\!\Gamma_L\!-\!\Gamma_R\! & 0\! \\
                         0\! & 0\! & \Gamma_L\! & -\Gamma_R\! \\
                       \end{array}
                     \right),
\end{equation}
where the tunneling rates between the DQD and the leads $\Gamma_\alpha(\epsilon)=2\pi\sum_k|t_{k,\alpha}|^2\delta(\epsilon-\epsilon_{k,\alpha})=\Gamma_\alpha$, $\alpha=L,R$, are assumed to be constant. A large bias across the DQD ensures that the (broadened) levels of the DQD are well inside the bias-window \cite{Gurvitz1996}. The inter-dot tunneling rates
\begin{equation}
\widetilde{\Gamma}_{\!\pm}(z)=T_c^2\left[\widetilde{g}_{\!+}(z_{\pm})+\widetilde{g}_{\!-}(z_{\mp})\right]
\end{equation}
are given by the bath-correlation functions $g_{\!\pm}(t)=e^{-\mathcal{F}(\mp t)}$, where $z_{\pm}=z\pm i\varepsilon+(\Gamma_L+\Gamma_R)/2$ and $\mathcal{F}(t)=\int_0^\infty d\omega \mathcal{J}(\omega)\{[1-\cos(\omega t)]\coth(\hbar\omega/2k_BT)+i\sin(\omega t)\}/\omega^2$. Here, $\mathcal{J}(\omega)=\sum_j|g_j|^2\delta(\omega-\omega_j)$ is the spectral function of the heat bath with temperature $T$. We consider an ohmic bath with $\mathcal{J}(\omega)=2\alpha\omega e^{-\omega/\omega_c}$, where $\alpha$ is the strength of the coupling to the DQD and $\omega_c$ is a high-frequency cut-off. For strong couplings to the heat bath, the inter-dot tunneling rates are valid to lowest order in $T_c^2$. The inhomogeneity reads $\tilde{\gamhat}(s,z)=-\widetilde{\W}(1,z)\rhohat^{\mathrm{stat}}/z$ \cite{Flindt08,Emary11b}. Strong Coulomb interactions between the quantum dots may be included by excluding the double-occupied state of the DQD.

\emph{Non-Markovian dephasing.}--- We evaluate the WTD in Laplace space using Eq.~(\ref{eq:WTDlap}) and find the analytic result
\begin{widetext}
\begin{equation}
\widetilde{\WTD}(z)=\frac{\Gamma_L\Gamma_R(z+\Gamma_L+\Gamma_R)^2\widetilde{\Gamma}_{\!+}(z)}{(z+\Gamma_L)(z+\Gamma_R)(\Gamma_L+\Gamma_R)[z\widetilde{\Gamma}_{\!-}(z)+\{z+\Gamma_L+\Gamma_R\}\{z+\widetilde{\Gamma}_{\!+}(z)\}]}.
\label{eq:WTDDQD}
\end{equation}
\end{widetext}
In general, the electron waiting time is determined in an interplay between the time-scales associated with the incoming electrons in the transport window and those associated with the nano-structure. The incoming electrons are on average separated by the mean waiting time $\bar{\tau}=h/eV$, where $V$ is the applied voltage \cite{Albert12}. In our example, the voltage is much larger than the energy-scales of the DQD, implying that the mean waiting time between the incoming electrons is much shorter than the time-scales of the DQD, and $\bar{\tau}$ does not appear above.

Focusing first on the uncoupled case ($\alpha=0$), we find that the WTD in the limit of vanishing tunneling rates ($\Gamma_L,\Gamma_R\rightarrow 0$) has imaginary poles at $z=\pm i \Delta$, where $\Delta=\sqrt{4T_c^2+\varepsilon^2}$ is the energy splitting of the hybridized states of the DQD, see also Ref.~\cite{Brandes08}. These poles correspond to coherent oscillations between the quantum dot levels with period $h/\Delta$.  This is clearly visible in Fig.~\ref{Fig2}a, showing the WTD in the time domain. The pole structures $\Gamma_{L/R}/(z+\Gamma_{L/R})$ in Eq.~(\ref{eq:WTDDQD}) are due to poissonian charge transfers between the DQD and the leads, which damp the oscillations. The coherent oscillations are gradually washed out as the temperature of the heat bath is increased and electrons are dephased as they tunnel through the DQD.

Next, we increase the coupling to the heat bath. Markovian dephasing in the weak-coupling limit has been discussed by Brandes \cite{Brandes08}. Here we can take these ideas further and examine the transition from Markovian to non-Markovian dephasing. In Fig.~\ref{Fig2}b, we present WTDs at zero temperature beyond the limit of weak system-bath coupling. As the coupling increases, the heat bath tends to localize electrons on the quantum dots and the inter-dot tunneling rate becomes suppressed. For large couplings, tunneling events are rare and uncorrelated and the transport process essentially becomes poissonian.

\begin{figure}
\includegraphics[width=0.9\columnwidth]{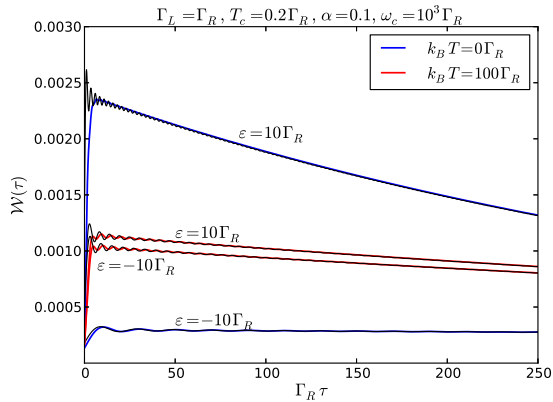}
\caption{(color online). Electronic waiting time distributions for dealigned levels. At low temperatures (blue lines), there is a strong asymmetry between positive and negative dealignments. The asymmetry diminishes at high temperatures (red lines). Red and blue curves are obtained from Eq.~(\ref{eq:WTDDQD}), while the black curves follow from the approximation in Eq.~(\ref{eq:WTDbottleck}).}
\label{Fig3}
\end{figure}

\emph{Heat bath \& WTD.}---  The DQD can be tuned to an interesting regime, where tunneling between the quantum dots becomes the rate-limiting step in the transport.  Choosing for instance the tunnel coupling $T_c$ or the dealignment of the quantum dot levels $\varepsilon$ such that $\widetilde{\Gamma}_{\!-}(z), \widetilde{\Gamma}_{\!+}(z)\ll \Gamma_L, \Gamma_R $, Equation (\ref{eq:WTDDQD}) reduces to
\begin{equation}
\widetilde{\WTD}(z)\simeq\frac{\widetilde{\Gamma}_{\!+}(z)}{z+\widetilde{\Gamma}_{\!+}(z)}.
\label{eq:WTDbottleck}
\end{equation}
This result offers the possibility of directly probing spectral properties of the heat bath (or another coupled conductor \cite{Braggio09}) through the detection of the electron waiting time, since the bath correlation functions enter the bath-assisted hopping rate $\widetilde{\Gamma}_{\!+}(z)$.  In Fig.~\ref{Fig3} we focus on the emission and absorption of energy to and from the heat bath as electrons tunnel from the left to the right quantum dot. At low temperatures, there is a clear asymmetry between the WTDs for positive and negative dealignments, since the heat bath mainly contributes to the transport for positive detunings by absorbing energy quanta from tunneling electrons. At high temperatures, this asymmetry disappears as the heat bath in addition can assist the tunneling process at negative detunings through the emission of energy quanta. Figure~3 shows that Eq.~(\ref{eq:WTDbottleck}) provides an excellent approximation to the exact results based on Eq.~(\ref{eq:WTDDQD}).

\emph{Conclusions}.--- We have presented a theory of electron waiting times for non-Markovian generalized master equations which unifies and generalizes a number of earlier approaches to waiting time distributions in the context of electronic transport. As an illustrative example we considered electron transport through a double quantum dot for which we examined non-Markovian dephasing mechanisms beyond the weak-coupling limit. We hope our method may pave the way for future investigations of memory effects and electron waiting times, similar to how full counting statistics and finite-frequency noise in non-Markovian quantum transport have been popular research topics in recent years.

\emph{Acknowledgments}.---  We thank M.~Albert, A.~Braggio, M.~B\"{u}ttiker, G.~Haack, and T.~Novotn\'{y} for useful comments on the manuscript. The work was supported by Swiss NSF.


%

\end{document}